\documentclass[12pt]{article}
\usepackage{amsmath, amssymb}
\usepackage{graphicx}
\usepackage{geometry}
\usepackage{color}
\usepackage{cite}
\newtheorem{remark}{Remark}

\title{Ramsey Approach to Quantum Mechanics}
\author{Edward Bormashenko\thanks{Corresponding author: edward@ariel.ac.il} \\ \small Ariel University, Engineering Sciences Faculty, Chemical Engineering Department, Ariel, Israel \\ Nir Shvalb \\ \small Department of Industrial Engineering and Management, Ariel University, Israel}
\date{}

\begin{document}

\maketitle

\begin{abstract}
Ramsey theory enables re-shaping of the basic ideas of quantum mechanics. Quantum observables represented by linear Hermitian operators are seen as the vertices of a graph. Relations of commutation define the coloring of edges linking the vertices: if the operators commute, they are connected with a red link; if they do not commute, they are connected with a green link. Thus, a bi-colored complete Ramsey graph emerges. According to Ramsey's theorem, a complete bi-colored graph built of six vertices will inevitably contain at least one monochromatic triangle; in other words, the Ramsey number \( R(3,3) = 6 \). In our interpretation, this triangle represents the triad of observables that could or could not be established simultaneously in a given quantum system. The Ramsey approach to quantum mechanics is illustrated with numerous examples, including the motion of a particle in a centrally symmetrical field.
\end{abstract}


\section{Introduction}

In this paper, we introduce the Ramsey approach for analyzing fundamental quantum mechanics problems. We implement graph theory for examining the foundations of quantum mechanics. In particular, we demonstrate that Ramsey theory enables the re-shaping of basic principles of quantum mechanics when operators/observables are seen as the vertices of a complete bi-colored graph, and their commutative properties represent the relations between the vertices.

In general, Ramsey theory addresses any mathematical problem that guarantees a structure of a given kind will contain a prescribed substructure. The classical problem in Ramsey theory is the so-called ``party problem," which asks the minimum number of guests (each of whom is either a ``friend" or a ``stranger" to the others) denoted \( R(m, n) \), such that at least \( m \) will know each other or at least \( n \) will not know each other \cite{ramsey2009, chartrand2021, graham1990, graham2015, landman2004, li2020, dinasso2019, katz2018, erdos1997, erdos1981, conlon2015}.

When Ramsey theory is re-shaped in the notions of graph theory, it states that any structure will necessarily contain an interconnected substructure \cite{graham1990,li2020}. The Ramsey theorem, in its graph-theoretic form, states that one will find monochromatic cliques in any edge color labelling of a sufficiently large complete graph \cite{graham1990,li2020}. Applications of Ramsey theory to physics remain scarce \cite{gaitan2012,bian2013,wouters2022}. In our paper, we address the Ramsey graphs emerging from Hermitian operators, representing the quantum mechanics observables \cite{messiah2014,landau1965,zettili2009,bohm1989}.

\section{Results and Discussion}

\subsection{Observables and Operators}

In quantum mechanics, an observable is a linear operator denoted as \( \hat{l} \) \cite{messiah2014, landau1965, zettili2009, bohm1989}. Observables manifest as Hermitian self-adjoint operators on a separable complex Hilbert space, representing the quantum state space, which possesses a complete orthogonal set of eigenfunctions \cite{messiah2014,landau1965,zettili2009,bohm1989}. All Hermitian operators do not possess a complete orthogonal set of eigenfunctions; however, the Hermitian operators capable of representing physical quantities possess such a set. To prove that a specific Hermitian operator is an observable is often a very difficult physical problem \cite{messiah2014}. The proof has already been given for simple cases, such as coordinate, momentum, and angular momentum \cite{messiah2014}. Two operators, \( \hat{f} \) and \( \hat{g} \), are said to commute with each other if the following equation holds:
\begin{equation}
\label{eq:commutator}
[\hat{f}, \hat{g}] = \hat{f} \hat{g} - \hat{g} \hat{f} = 0
\end{equation}
If a particle can be in a definite state for two observables, then the two operators associated with those observables will commute \cite{messiah2014,landau1965,zettili2009,bohm1989}. The converse is therefore also true: if two operators do not commute, then it is not possible for a quantum state to have a definite value of the corresponding two observables at the same time \cite{messiah2014,landau1965,zettili2009,bohm1989}. The operator of momentum is defined as $\hat{p} = -j\hbar\nabla$
or in components: 
\begin{equation}
\label{eq:momentum}
p_x = -j\hbar \frac{\partial}{\partial x}, \quad p_y = -j\hbar \frac{\partial}{\partial y}, \quad p_z = -j\hbar \frac{\partial}{\partial z}
\end{equation}
The operator corresponding to the coordinate \( q \) is simply multiplication by \( q \). The spectrum of this operator is continuous. The commutation rules for \( \hat{p} \) and \( x \) are given by:
\begin{equation}
\label{eq:p_x_commutator}
[p_i, x_k] = -j\hbar \delta_{ik},
\end{equation}
where \( \delta_{ik} \) is the Kronecker delta. Eq. \ref{eq:p_x_commutator} demonstrates that the coordinate of the particle along one of the axes can have a definite value at the same time as the components of the momentum along the other two axes; Conversely, the position and momentum components along the same axis cannot be simultaneously determined with arbitrary precision. For the angular momentum component operators of a particle, denoted

For the angular momentum component operators of a particle, denoted \( \hat{l}_i \) for \( i=1,2,3 \), we have:
\begin{equation}
\label{eq:angular_momentum}
\hbar \hat{l}_x = y p_z - z p_y; \quad \hbar \hat{l}_y = z p_x - x p_z; \quad \hbar \hat{l}_z = x p_y - y p_x
\end{equation}
The rules for commutation of the angular momentum operators with the operators of coordinates and linear momenta are:
\begin{equation}
\label{eq:l_x_commutators}
[\hat{l}_i, x_k] = i \epsilon_{ikl} x_l
\end{equation}
\begin{equation}
\label{eq:l_p_commutators}
 [\hat{l}_i, p_k] = i \epsilon_{ikl} p_l
\end{equation}
where $\epsilon_{ikl}$ is the antisymmetric unit tensor of rank three with summation implied by Einstein's summation convention. The rules of commutation for the operator of angular momentum are given by 
\begin{equation}
\label{eq:l_l_commutators}
[\hat{l}_i, \hat{l}_k] = i \epsilon_{ikl} \hat{l}_l
\end{equation}
For the commutation rules established for the total angular momentum of the system, denoted \( \hat{L} \), see ref. \cite{landau1965}. It should be emphasized that the commutation relations given by Eqs. \ref{eq:l_l_commutators} and \ref{eq:p_x_commutator} are non-transitive. The transitive Ramsey numbers are different from the non-transitive ones \cite{choudum1999}.

\subsection{Observables as a complete graph $K_n$}

\begin{figure}[ht!]
    \centering
\includegraphics[width=\textwidth]{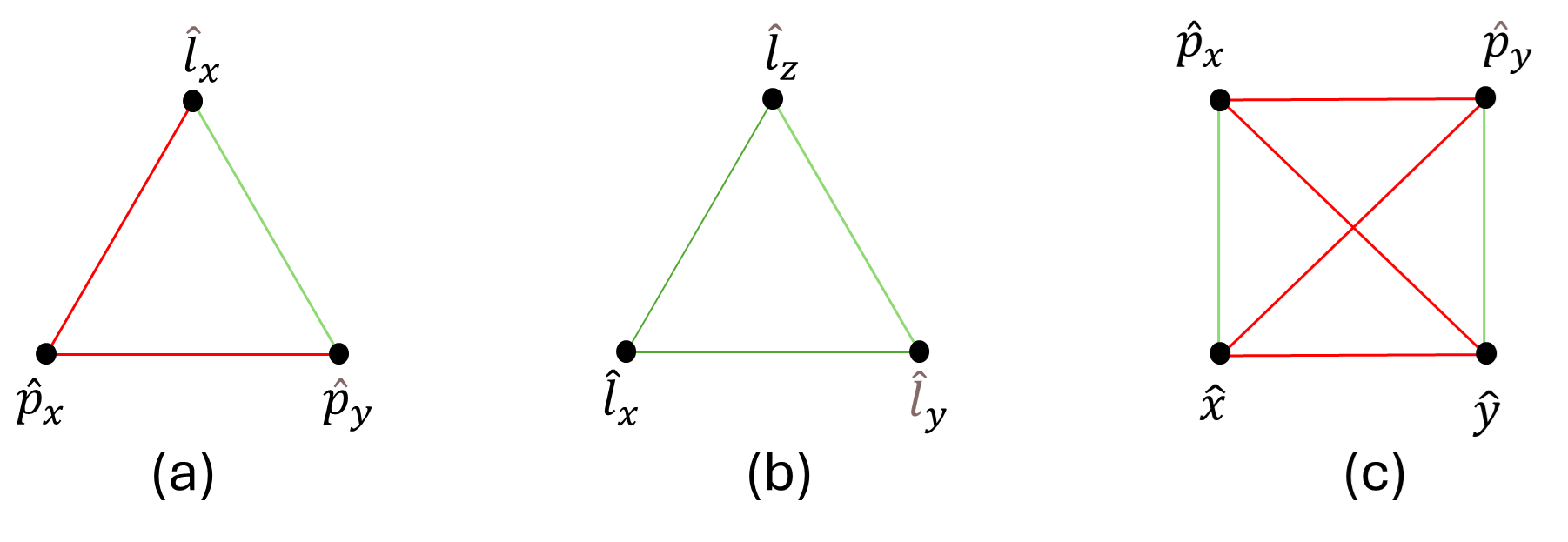}
    \caption{Graphs representing quantum observables with three and four vertices are depicted. (a) The observables are $\hat{p}_x$, $\hat{p}_y$, and $\hat{l}_x$. The graph is bi-colored, reflecting the commutation relations between the observables. (b) The observables are $\hat{l}_x$, $\hat{l}_y$, and $\hat{l}_z$. The graph is mono-colored (green), indicating that none of the observables commute. (c) A complete bi-colored graph is shown for the four observables $\hat{p}_x$, $\hat{p}_y$, $\hat{x}$, and $\hat{y}$. No monochromatic triangle is present in this graph, meaning no set of three observables commute or all fail to commute simultaneously.}
\label{fig:simple_examples}
\end{figure}

\begin{figure}[ht!]
\centering
\includegraphics[width=\textwidth]{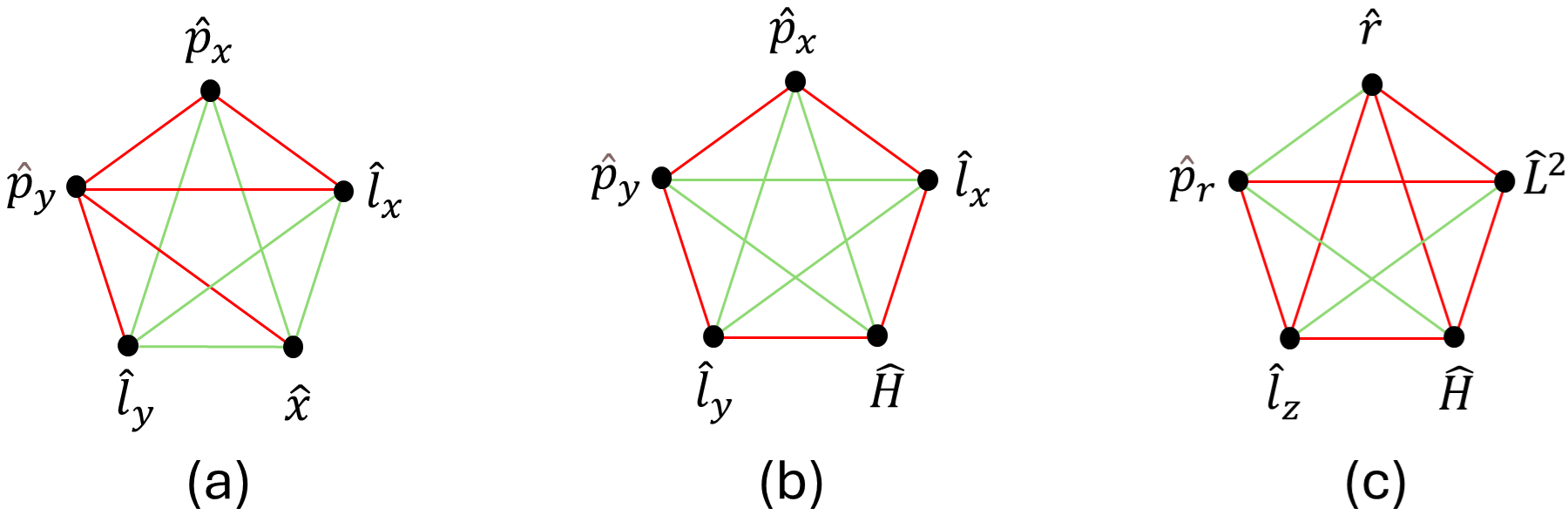}
\caption{(a) Graph containing five vertices, namely $\hat{p}_x$, $\hat{p}_y$, $\hat{l}_x$, $\hat{l}_y$, and $\hat{x}$. The triangle $(\hat{p}_x,\hat{l}_x,\hat{x})$ is monochromatic green.
(b) Complete, bi-colored graph containing five vertices, namely $\hat{p}_x$, $\hat{p}_y$, $\hat{l}_x$, $\hat{l}_y$, and $\hat{H}$. The graph does not contain any monochromatic triangle.
(c) Bi-colored, complete graph arising from the operators $\hat{r}$, $\hat{p}_r$, $\hat{l}_z$, $\hat{H}$, and $\hat{L}^2$, where $\hat{L}^2 = \hat{l}_x^2 + \hat{l}_y^2 + \hat{l}_z^2$. The triangles $(\hat{r},\hat{p}_r,\hat{L}^2)$ and $(\hat{p}_r,\hat{l}_z,\hat{L}^2)$ are monochromatic red, and the triangle $(\hat{r},\hat{l}_x,\hat{p}_r)$ is monochromatic green.}
\label{fig:penthagons_2}
\end{figure}

We start with the motion of a single quantum particle $\mu_A$. Following mathematical procedure enabling converting of the observables into graph is suggested:  observables themselves are represented by the vertices of the graph. The relation between observables/vertices/operators are established by the commutation rules, namely: if the observables/vertices commute, they are linked with the red edge (they are “friends” within the terminology of the Ramsey theory). And, when the observables/vertices do not commute (thus, they are “strangers”), they are connected with the green link, as shown in Figure \ref{fig:simple_examples}.     
In Figure \ref{fig:simple_examples}, the vertices in inset (a) form a bi-colored graph (see Eq. \ref{eq:l_p_commutators}), while Figure \ref{fig:penthagons_2}(b)  forms a mono-colored graph (see Eq. \ref{eq:l_l_commutators}), showing the angular momentum components cannot be measured simultaneously \cite{messiah2014,landau1965,zettili2009,bohm1989}, the vertices in Figure \ref{fig:penthagons_2}(c) shows a complete graph with four vertices, where no monochromatic triangle is present, meaning no triad of observables can be measured simultaneously \cite{graham1990}.
\begin{figure}[ht!]
\centering
\includegraphics[width=1\textwidth]{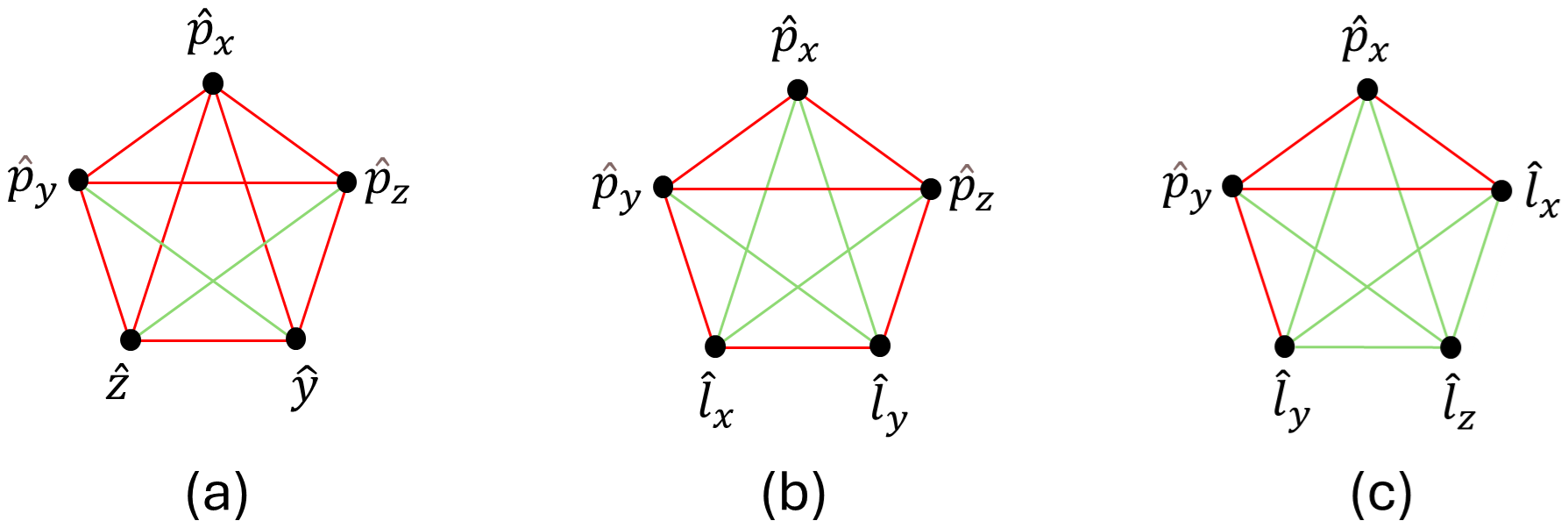}
\caption{(a) Bi-colored graph emerging from the observables: $\hat{p}_x$, $\hat{p}_y$, $\hat{p}_z$, $\hat{z}$, and $\hat{y}$. The triangles $\hat{p}_x\hat{p}_y\hat{p}_z$, $\hat{p}_x\hat{p}_y\hat{z}$, $\hat{p}_x\hat{p}_z\hat{y}$, and $\hat{p}_y\hat{z}\hat{y}$ are monochromatic red. (b) Bi-colored, complete graph emerging from the observables: $\hat{p}_x$, $\hat{p}_y$, $\hat{p}_z$, $\hat{l}_y$, and $\hat{l}_z$. The triangle $(\hat{p}_x,\hat{p}_y,\hat{l}_y)$ is monochromatic red, and the triangle $(\hat{p}_x,\hat{p}_z,\hat{l}_z)$ is monochromatic.
(c) The graph arising from the observables $\hat{p}_x$, $\hat{p}_y$, $\hat{l}_x$, $\hat{l}_y$, and $\hat{l}_z$. The triangles $(\hat{p}_x,\hat{l}_x,\hat{l}_y)$ and $(\hat{l}_x,\hat{l}_y,\hat{l}_z)$ are monochromatic green.}
\label{fig:penthagons_1}
\end{figure}
The inset (a) of Figure \ref{fig:penthagons_2} presents the bi-colored, complete Ramsey graph containing five vertices, namely $\hat{p}_x$, $\hat{p}_y$, $\hat{l}_x$, $\hat{l}_y$, and $\hat{x}$. The triangle $(\hat{p}_x,\hat{l}_y,\hat{x})$ is monochromatic green, indicating that the observables $\hat{p}_x$, $\hat{l}_y$, and $\hat{x}$ cannot be established simultaneously. A monochromatic quadrangle, $(\hat{p}_x,\hat{p}_y,\hat{l}_x\hat{x})$, is also recognized in Figure \ref{fig:penthagons_2}(a). However, this does not imply that all these observables can be measured simultaneously, as pairs such as $(\hat{p}_x, \hat{x})$ and $(\hat{p}_y, \hat{l}_x)$ cannot be measured at the same time. Although graphs may contain mono-colored triangles, it is possible to construct a bi-colored complete graph that does not contain any mono-colored triangle. This is related to the Ramsey number $R(3,3) = 6$, as demonstrated in inset Figure \ref{fig:penthagons_2}(b).

In inset Figure \ref{fig:penthagons_2}(b), the graph represents the observables $\hat{p}_x$, $\hat{p}_y$, $\hat{l}_x$, $\hat{l}_y$, and $\hat{H}$, where $\hat{H}$ is the Hamiltonian of the particle. Assuming that the Hamiltonian $\hat{H}$ explicitly depends on the particle's coordinates, it does not commute with the operators $\hat{p}_x$, $\hat{p}_y$, $\hat{l}_x$, or $\hat{l}_y$ \cite{messiah2014,landau1965,zettili2009,bohm1989}. As a result, the graph in inset Figure \ref{fig:penthagons_2}(b) does not contain any monochromatic triangle, illustrating Ramsey's theorem, $R(3,3) = 6$.

To further illustrate these ideas, we consider the motion of a particle $\mu_A$ in a centrally symmetric field $U(r)$ \cite{messiah2014,landau1965,zettili2009,bohm1989}. This scenario involves two particles, $m_1$ and $m_2$, whose motion can be described by the wave function $\psi(\vec{r}_1, \vec{r}_2) = \phi(\vec{R})\psi(\vec{r})$, where $\phi(\vec{R})$ describes the motion of the center of mass (as a free particle with mass $m_1 + m_2$), and $\psi(\vec{r})$ describes the relative motion of the particles, with an effective mass $m = \frac{m_1 m_2}{m_1 + m_2}$ moving in the centrally symmetric field $U(r)$ \cite{messiah2014,landau1965,zettili2009,bohm1989}. The Schr\"odinger equation for the particle $m$ in the field $U(r)$ is given by:
\begin{equation}
\label{eq:Schrodinger}
\Delta \psi + \frac{2m}{\hbar^2} [E - U(r)]\psi = 0, \tag{8}
\end{equation}
where $E$ is the particle's energy. This reduces the problem to the motion of a single particle in the centrally symmetric field $U(r)$ \cite{messiah2014,landau1965,zettili2009,bohm1989}. The graph arising from the operators $\hat{r}$, $\hat{p}_r$, $\hat{l}_x$, $\hat{H}$, and $\hat{L}^2$, where $\hat{L}^2 = \hat{l}_x^2 + \hat{l}_y^2 + \hat{l}_z^2$, is shown in Figure \ref{fig:penthagons_2}(c). For commutation rules, see refs. \cite{messiah2014,landau1965,zettili2009,bohm1989}. The triangle $(\hat{p}_r,\hat{l}_z,\hat{L}^2)$ is monochromatic red, while the triangle $(\hat{r},\hat{p}_r,\hat{H})$ is monochromatic green. This indicates that the observables $\hat{p}_r$, $\hat{l}_z$, and $\hat{L}^2$ can be established simultaneously, whereas $\hat{r}$, $\hat{p}_r$, and $\hat{H}$ cannot. Additionally, the triangle $(\hat{r},\hat{L}^2,\hat{l}_z)$ is monochromatic red, meaning this triad of observables can also be measured simultaneously.

We recognize that in Figure \ref{fig:penthagons_2}(c), the triangle formed by the observables $(\hat{p}_r, \hat{l}_z, \hat{L}^2)$ is monochromatic red, while the triangle formed by $(\hat{r}, \hat{p}_r, \hat{H})$ is monochromatic green. This means that the triad of observables $(\hat{p}_r, \hat{l}_z, \hat{L}^2)$ can be established simultaneously, whereas the triad $(\hat{r}, \hat{p}_r, \hat{H})$ cannot be established at the same time. Additionally, the triangle formed by $(\hat{r}, \hat{L}^2, \hat{l}_z)$ is monochromatic red, indicating that this set of observables can also be measured simultaneously.

\subsection{Quantum observables graph $K_6$}

\begin{figure}[ht!]
\centering
\includegraphics[width=0.3\textwidth]{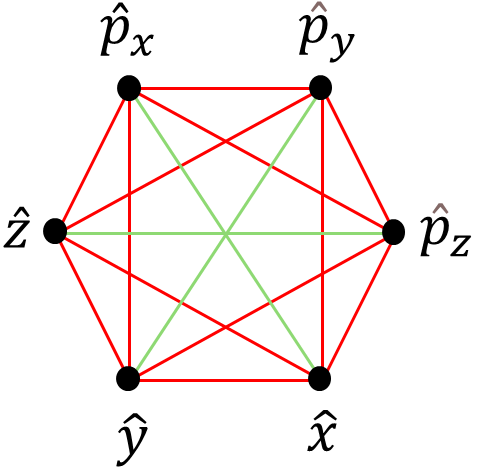}
\caption{Complete bi-colored graph emerging from the quantum observables $\hat{p}_x$, $\hat{p}_y$, $\hat{p}_z$, $\hat{x}$, $\hat{y}$, and $\hat{z}$. The triangles formed by $(\hat{p}_x, \hat{p}_y, \hat{p}_z)$, $(\hat{p}_y, \hat{p}_z, \hat{x})$, $(\hat{p}_z, \hat{x}, \hat{y})$, $(\hat{x}, \hat{y}, \hat{z})$, and others are monochromatic red. No monochromatic green triangle is recognized.}
\label{fig:hexagon}
\end{figure}

Now consider the complete, bi-colored graph with six vertices, representing six quantum observables/Hermitian operators, as depicted in Figure \ref{fig:hexagon}. This graph reflects the 3D motion of a quantum particle $\mu_A$. The observables are $\hat{p}_x$, $\hat{p}_y$, $\hat{p}_z$, $\hat{x}$, $\hat{y}$, and $\hat{z}$. The triangles $(A,B,C)$ formed by $(\hat{p}_x, \hat{p}_y, \hat{p}_z)$, $(\hat{p}_y, \hat{p}_z, \hat{x})$, $(\hat{p}_z, \hat{x}, \hat{y})$, $(\hat{p}_x, \hat{y}, \hat{z})$, and $(\hat{x}, \hat{y}, \hat{z})$ are all monochromatic red. This means that the corresponding triads of observables, such as $(\hat{p}_x, \hat{p}_y, \hat{p}_z)$, $(\hat{p}_y, \hat{p}_z, \hat{x})$, and $(\hat{x}, \hat{y}, \hat{z})$, can be established simultaneously. No green monochromatic triangle is recognized in the graph, indicating that no triad of observables can be established simultaneously.

The reported result is far from trivial. Indeed, any graph constructed with six vertices/observables/Hermitian operators will necessarily contain at least one monochromatic triangle, regardless of the specific observables involved. This conclusion arises from the Ramsey analysis of the observables themselves. It is noteworthy that the vertices of the graph may represent a complete set of commuting observables (CSCO) \cite{messiah2014,landau1965,zettili2009,bohm1989}. What happens when the number of vertices in the graph increases? We know only a limited number of Ramsey numbers, and there is no known algorithm for their exact calculation \cite{graham1990,li2020}. The problem of calculating Ramsey numbers remains unsolved \cite{graham1990,li2020}.

\subsection{Quantum observables hypergraph $K_3^{(10)}$}

Consider a set of quantum operators $\{A_1, A_2, \dots, A_n\}$, where $n \geq 10$, acting on a Hilbert space. Suppose there exists at least one pair of operators $A_{i_1}$ and $A_{i_2}$ such that \hbox{$[A_{i_1}, A_{i_2}] \neq 0$}, indicating that not all operators in the set commute. This implies that the algebra is non-trivial, with non-zero commutation relations between some operators.

To explore the structure of this set, we define a hypergraph $K_3^{(n)}$, where each vertex corresponds to an operator $A_i$, and a hyperedge (or triangle) connects some three vertices $A_i$, $A_j$, and $A_k$ if these operators satisfy the Jacobi identity:
\begin{equation}
\label{eq:jacobi}
[A_i, [A_j, A_k]] + [A_j, [A_k, A_i]] + [A_k, [A_i, A_j]] = 0.
\end{equation}
By applying principles from Ramsey theory \cite{conlon2010hypergraph}, we can assert that for sufficiently large $n$, such as $n \geq 10$, there must exist at least one triplet of operators $(A_i, A_j, A_k)$ for which the Jacobi identity holds. 

The Jacobi identity plays a vital role in ensuring the coherence of symmetries and conservation laws in quantum systems. For instance, when considering the angular momentum components $(J_x, J_y, J_z)$, the Jacobi identity ensures that their commutation relations remain consistent, preserving the conservation of angular momentum.
Consider, for example, the set $\{\hat{H}, \hat{L}^2, \hat{p}_x, \hat{p}_y, \hat{p}_z, \hat{x}, \hat{y}, \hat{z}, \hat{l}_x, \hat{l}_y, \hat{l}_z\}$ for a single particle $\mu_A$ which form the basis of the quantum mechanical description of a single particle moving in three-dimensional space.  
For most standard quantum systems, the Jacobi identity will hold for all possible triads of these operators. This is due to the well-defined commutation relations between position, momentum, angular momentum, and the Hamiltonian. The position and momentum operators, as well as the angular momentum components, respect the Jacobi identity because they satisfy the canonical commutation relations of quantum mechanics. Similarly, combinations of angular momentum, position, and momentum operators maintain the consistency of these relations. In the case of a Hamiltonian of the form $\hat{H} = \frac{\hat{p}^2}{2m} + V(\hat{x})$, the Jacobi identity will also hold for combinations of $\hat{H}$ with momentum and position operators. However, for more complex Hamiltonians, such as those with interactions or external fields, the commutation relations involving $\hat{H}$ may become more intricate. 

\begin{remark}
 In gauge theories the identity ensures that the interactions between gauge fields, like the electromagnetic field, are coherent, thereby preserving charge conservation.
\end{remark}

\subsection{Quantum observables hypergraph $K_3^{(10)}$ of a particle system}

When considering two distinct particles $\mu_A$ and $\mu_B$, each has its own set of quantum operators. For a non-entangled system, the operators for each particle act independently. The total Hilbert space is a tensor product, and the operators commute across particles.  In the case of entanglement, the quantum state is no longer separable and takes the form of a superposition:
\begin{equation}
|\psi\rangle = \sum_{i,j} c_{ij} |\psi_A^i\rangle |\psi_B^j\rangle,
\end{equation}
where $c_{ij}$ are complex coefficients. Entanglement has several key impacts on the behavior of operators for two distinct particles $\mu_A$ and $\mu_B$. First, non-local correlations arise, meaning that measurements on particle $\mu_A$ can affect the outcomes for particle $\mu_B$, even if they are spatially separated. Second, while the Jacobi identity holds locally for each particle, combined operators, such as joint observables, may exhibit non-trivial correlations due to the entanglement. 

\subsection{Interference Patterns in Multi-Slit Experiments}

In quantum interference experiments, the complexity of the interference pattern increases as the number of slits rises from one to five. For a particle passing through multiple slits, the interference pattern is the result of constructive and destructive interference between the different possible paths. The probability distribution of detecting the particle on the screen is given by the squared modulus of the superposition of the particle's wavefunction at each slit:
\begin{equation}
P(\mathbf{r}) = \left| \psi_1(\mathbf{r}) + \psi_2(\mathbf{r}) + \dots + \psi_n(\mathbf{r}) \right|^2,
\end{equation}
where $n$ is the number of slits, and $\psi_i(\mathbf{r})$ represents the wavefunction associated with the particle passing through slit $i$. The more slits involved, the more intricate the interference pattern due to the increased number of terms contributing to the superposition.

When entanglement between the particle and another quantum system (such as a measurement device or another particle) is introduced, the coherence between the possible paths is reduced. This decoherence can be modeled by a reduced density matrix formalism, where tracing out the entangled system introduces a decoherence factor $\gamma$ \cite{nielsen2010quantum}:
\begin{equation}
\rho_{\text{reduced}} = \text{Tr}_{\text{env}} (\rho_{\text{total}}) = \sum_i p_i \rho_i e^{-\gamma t}.
\end{equation}
\subsection{Conjecture}
As the entanglement increases, the interference pattern on the screen is suppressed, leading to a blurring of the interference fringes. In particular, in a 1- to 5-slit experiment, the increasing number of possible paths leads to increasingly complex interference patterns, but entanglement gradually destroys these patterns by reducing the coherence between the paths.

However, the authors suspect that the situation becomes significantly more intriguing when considering a six-slit experiment. According to Ramsey theory, a complete graph constructed from six quantum observables (representing slit paths) will necessarily contain at least one monochromatic triangle due to the Ramsey number $R(3,3)$. This implies that certain sets of slit paths will evidently be highly correlated or entangled. In the context of the interference pattern, this suggests that instead of a gradual loss of coherence, the interference may exhibit unique characteristics—specific sets of paths may either show enhanced interference or be entirely suppressed due to the presence of non-local correlations between the paths.

The total interference pattern in the presence of entanglement can be described by a modified superposition that incorporates entanglement-induced correlations:
\begin{equation}
P_{\text{entangled}}(\mathbf{r}) = \left| \sum_i \psi_i(\mathbf{r}) + \sum_{i,j} C_{ij} \psi_i(\mathbf{r}) \psi_j(\mathbf{r}) \right|^2,
\end{equation}
where $C_{ij}$ represents the correlation between the paths $i$ and $j$ induced by entanglement. This correlation term modifies the interference pattern, leading to novel effects not present in the non-entangled case. To summerize while entanglement leads to a gradual decoherence and destruction of the interference pattern in the 1- to 5-slit experiments, in the six-slit experiment, the presence of a monochromatic triangle in the associated Ramsey graph implies that certain sets of slit paths are strongly correlated. As a result, the six-slit experiment is expected to exhibit unique interference effects, with some paths showing stronger interference or novel correlations due to the underlying graph structure, potentially leading to enhanced or suppressed fringes.
\bibliographystyle{plain}
\bibliography{main}
\end{document}